\documentclass[twocolumn,showpacs,amsmath,amssymb]{revtex4}

\usepackage{graphicx}
\usepackage{array}
\usepackage{longtable}
\usepackage{rotating,booktabs}
\usepackage{booktabs,threeparttable}

\begin{document}

\title{Dynamic polarizabilities for the low lying states of Ca$^+$}

\author{Yong-Bo Tang$^{1,2}$, Hao-Xue Qiao$^{1}$, Ting-Yun Shi$^{2}$ and J. Mitroy$^{3}$}

\affiliation{$^{1}$Department of Physics, Wuhan University, Wuhan
430072, P. R. China}

\affiliation {$^{2}$State Key Laboratory of Magnetic Resonance and
Atomic and Molecular Physics, Wuhan Institute of Physics and
Mathematics, Chinese Academy of Sciences, Wuhan 430071, P. R. China}

\email{jxm107@physics.anu.edu.au} \affiliation {$^{3}$School of Engineering,
Charles Darwin University, Darwin NT 0909, Australia}

\date{\today}

\begin{abstract}

The dynamic polarizabilities of the $4s$, $3d$ and $4p$ states of Ca$^+$,
are calculated using a relativistic structure model.  The wavelengths
at which the Stark shifts between different pairs of transitions are zero
are computed.  Experimental determination of the magic wavelengths
can be used to estimate the ratio of the $f_{3d_{J}\!\to \!4p_{J'}}$ and
$f_{4s_{1/2} \! \to \! 4p_{J'}}$ oscillator strengths.   This
could prove valuable in developing better atomic structure models and in
particular lead to improved values of the polarizabilities needed in the
evaluation of the blackbody radiation shift of the Ca$^+$ ion.

\end{abstract}

 \maketitle

\section{Introduction}

The dynamic polarizability of an atom or ion gives a measure of the energy
shift of the atom or ion when immersed in an electromagnetic field
\cite{miller77a,bonin97a,mitroy10a}.
For any given state, one can write
\begin{equation}
\Delta E = - \frac{1}{2} \alpha_d(\omega) F^2 \ ,
\end{equation}
where $\alpha_d(\omega)$ is the polarizability of the quantum state at frequency
$\omega$, and $F$ is a measure of the strength of the AC electromagnetic field.
The value of the dynamic polarizability in the $\omega \to 0$ limit is
the static dipole polarizability.

The magic wavelength for a transition is the wavelength for which the
AC Stark shift of the transition energy is zero
\cite{katori03a,santra04a,arora07c,safronova12c}.
The identification of magic wavelengths and their use in making
optical lattices has resulted in the development of optical lattice
clocks which have the potential to exceed the performance characteristics
of the existing standard for time, namely the cesium microwave clock
\cite{bauch03a,takamoto03a,gill05a,lorini08a,gill11a}.

However, the experimental determination of magic wavelengths also provides
valuable information to constrain the atomic structure models that
are used to estimate the impact of Stark shifts on the performance on
atomic and ion clocks \cite{porsev08a,herold12a}.  A parameter related to
the magic wavelength is the tune-out wavelength.  The
tune-out wavelengths for an atomic state are the wavelengths at which the
polarizability for that state goes to zero \cite{leblanc07a,arora11a,holmgren12a}.
It should be noted that most atomic states have a number of tune-out
wavelengths just like most atomic transitions have a variety
of magic wavelengths.

The advantage of magic and tune-out wavelength measurements  are
that they are effectively null experiments.  They measure the frequencies
at which polarizability related quantities are equal to zero.  Therefore
they do not rely on a precise determination of the strength of a static
electric field or the intensity of a laser field.  This makes it possible
to determine the magic wavelengths to a high degree of precision
\cite{takamoto03a,mckeever03a,degenhardt04a,kien05a,barber08a,yi12a}.

There have been a number of theoretical studies of the properties of the low-lying Ca$^+$
ion \cite{kreuter05a,mitroy08b,jiang08a,sahoo09a,arora07a,safronova10e,safronova11a}
by 3 different research groups. One of these groups \cite{mitroy08b} used a
non-relativistic approach while the other two groups used explicitly relativistic
formulations \cite{sahoo09a,arora07a,safronova10e,safronova11a}.  One of the
singular features about the relativistic calculations are significant
differences between predictions of the properties of spin-orbit doublets.  The
relativistic all-order many body perturbation theory method predicts relatively
small non-geometric differences between the line strengths of the $3d_J$ and $4p_J$
spin-orbit doublets \cite{kreuter05a,safronova11a}.  The relativistic coupled
cluster approach typically gives much larger differences~\cite{sahoo09a}.
One of the secondary aims of the present work is to shed light on these
differences.

The present manuscript reports calculations of the dynamic polarizabilities of
the five lowest states of Ca$^+$.  The Hamiltonian used is a fully relativistic
version of a semi-empirical fixed core potential that has been successfully
applied to the description of many one and two electron atoms
\cite{mitroy88d,mitroy03f,mitroy09a,mitroy09b}.  While there are many
differences in the technical detail, the underlying philosophy and the
effective Hamiltonian for the valence electron are essentially the
same once the relativistic modifications are taken into account.
Magic wavelengths for the $4s \to 3d_{3/2,5/2}$ and $4s \to 4p_{1/2,3/2}$
transitions are given.  The dynamic polarizability of the ground
Ca$^+$($4s$) state is dominated by the $4s \to 4p_J$ transitions and
its accuracy is largely dependent on the accuracy of the transition
matrix elements connecting the $4s$ and $4p_J$ states.
The description of transitions involving the Ca$^+$($3d$) state is
complicated by the effect that the $3d$ electrons have on the core electrons.  The $3d$
orbitals have the smallest $\langle r \rangle$ expectation values
of any of the valence electrons and this does distort the wavefunctions
for the outermost core electrons \cite{vaeck92a,mitroy93a}.  One consequence
of this are greater uncertainties in the calculation of transition matrix
elements involving the $3d_J$ states \cite{mitroy93a,arora07a,safronova11a}

All results reported in this paper are given in atomic units with the
exception of the lifetimes which are given in seconds.  The value adopted
for the speed of light is $c = 137.035 999 074(44)$ a.u.

\section{Formulation and energies}

\subsection{Solution of the Dirac-Fock equation for closed shell atomic system }

The calculation methodology is as follows.  The first step involves
a Dirac-Fock (DF) calculation of the Ca$^{2+}$ ground state.
The DF calculation begins with the equation
\begin{equation}
\left( \sum_{i}^{N}{H_{Di}}+\sum_{i<j}^{N}{\frac{1}{r_{ij}}} \right) \psi(r)=E\psi(r),
\end{equation}
where $H_{Di}$ is the single-electron Dirac Hamiltonian
\begin{equation}
H_{Di}^{}=c \alpha_{i}\cdot{p_{i}}+c^{2}(\beta_{i}-1)+V(r_{i}).
\end{equation}
where $c$ is the speed of light, ${\mathbf p}$ is the momentum operator,
and $\alpha$ and $\beta$ are the Dirac matrices ~\cite{grant07a}.

The orbitals of the DF wave function, $\psi(r)$, can be written as
\begin{eqnarray}
\psi(r)=\frac{1}{r}\left(\begin{array}{c}
 g_{n\kappa}(r)\Omega_{\kappa m}(\hat{r})\\
if_{n\kappa}(r)\Omega_{-\kappa m}(\hat{r})\\
\end{array}
\right),
\end{eqnarray}
where $g_{n\kappa}(r)$ and $f_{n\kappa}(r)$ are the large and small
components, $\Omega_{\kappa m}(\hat{r})$ and
$\Omega_{-\kappa m}(\hat{r})$ correspond to the angular components.
The radial Dirac equation for an orbital can be expressed schematically
as

\begin{eqnarray}
\left (
\begin{array}{cc}
V(r)+V_{\rm DF}(r)                             & -c(\frac{d}{dr}-\frac{\kappa}{r})\\
c(\frac{d}{dr}+\frac{\kappa}{r})  &-2c^{2}+V(r)+V_{\rm DF}(r)\\
\end{array}
\right ) \left (
\begin{array}{c}
g_{n\kappa}(r)\\
f_{n\kappa}(r)\\
\end{array}
\right ) =\varepsilon \left (
\begin{array}{c}
g_{n\kappa}(r)\\
f_{n\kappa}(r)\\
\end{array}
\right ),
\end{eqnarray}
where $V_{\rm DF}$ is called the Dirac-Fock potential, and $V(r)$ is the
interaction potential between the electron and the nucleus. A Fermi nuclear
distribution approximation is usually adopted for many-electron atomic system.

The single particle orbitals are written as linear combinations of analytic
basis functions and so the method of Roothaan \cite{roothaan51a,mitroy99f}
is used to recast the DF equations into a set of matrix equations.  The functions
chosen are B-splines with Notre-Dame boundary conditions \cite{johnson88a}.
The large  and small components are expanded in terms of
a B-spline basis of $k$ order defined on the finite cavity $[0,R_{\rm max}]$,
\begin{equation}
g_{n\kappa}(r)=\sum_{i=1}^{N}{C_{i}^{g,n}B_{i,k}(r)}
\end{equation}
\begin{equation}
f_{n\kappa}(r)=\sum_{i=1}^{N}{C_{i}^{f,n}B_{i,k}(r)}.
\end{equation}
The finite cavity is set as a knots sequence, ${\{t_{i}\}}$,
satisfying an exponential distribution \cite{tang12b,tang13c}.
The specifics of the grid were that $R_{\rm max}=60$ $a_0$ and
50 B-splines of order $k = 7$ were used to represent the
single particle states.
Using the Galerkin
method and MIT-bag-model boundary conditions~\cite{johnson88a}, the DF equations
were solve by iteration until self-consistency was achieved. The single-electron
orbital (Koopmans) energies of the closed shell Ca$^{2+}$ ion agreed with those
computed with the GRASP92 program
\cite{parpia96a} to better than 10$^{-5}$ a.u.

\begin{table}
\caption{\label{tab1} Theoretical and experimental energy levels (in Hartree)
for some of the low-lying states of Ca$^+$. The energies are given relative
to the energy of the Ca$^{2+}$ core. The experimental data were taken from
the NIST tabulation \cite{nistasd500}.
 }
\begin{ruledtabular}
\begin {tabular}{lccc}
\multicolumn{1}{c}{Level}&\multicolumn{1}{c}{DF}&\multicolumn{1}{c}{DFCP}&\multicolumn{1}{c}{Experiment~\cite{nistasd500}}\\
\hline
$4s_{1/2}$ &-0.4166315  &-0.4362777    &-0.4362776    \\
$3d_{3/2}$ &-0.3308695  &-0.3740834    &-0.3740827    \\
$3d_{5/2}$ &-0.3307597  &-0.3738074    &-0.3738062    \\
$4p_{1/2}$ &-0.3099986  &-0.3214966    &-0.3214966    \\
$4p_{3/2}$ &-0.3090889  &-0.3204818    &-0.3204810    \\
$5s_{1/2}$ &-0.1933158  &-0.1983486    &-0.1985876    \\
$4d_{3/2}$ &-0.1687383  &-0.1751536    &-0.1772989    \\
$4d_{5/2}$ &-0.1686641  &-0.1750622    &-0.1772114    \\
$5p_{1/2}$ &-0.1567656  &-0.1603178    &-0.1604688    \\
$5p_{3/2}$ &-0.1564329  &-0.1600612    &-0.1601123    \\
\end{tabular}
\end{ruledtabular}
\end{table}

\subsection{Polarization potential}

The effective potential of the valence electron with the core is then
written
\begin{eqnarray}
V_{\rm core}  &=&   V_{\rm dir}({\bf r}) + V_{\rm exc}({\bf r})
+  V_{\rm pol}({\bf r}) \ .
\end{eqnarray}
The direct and exchange interactions of the valence electron with
the DF core were calculated exactly.  The $\ell$-dependent polarization
potential, $V_{\rm pol}$, was semi-empirical in nature with the
functional form
\begin{equation}
V_{\rm pol}(r)  =  -\sum_{\ell j} \frac{\alpha_{\rm core} g_{\ell j}^2(r)}{2 r^4}
 |\ell j \rangle \langle \ell j| .
 \label{polar1}
\end{equation}
The coefficient, $\alpha_{\rm core}$
is the static dipole polarizability of the core and
$g_{\ell j}^2(r) = 1-\exp\bigl(-r^6/\rho_{\ell,j}^6 \bigr)$
is a cutoff function designed to make the polarization potential
finite at the origin.  The static dipole polarizability
core was set to
$\alpha_{\rm core} = 3.26$ a.u. ~\cite{safronova11a}.
The cutoff parameters, $\rho_{\ell,j}$ were
tuned to reproduce the binding energies of the $ns$
ground state and the $np_J$, $nd_J$ excited states.
Values of the cutoff parameters are $\rho_{0,{1/2}} = 1.7419$ $a_0$,
$\rho_{1,{1/2}} = 1.6389$ $a_0$,  $\rho_{1,{1/2}} = 1.6354$ $a_0$,
$\rho_{2,{3/2}} = 1.8472$ $a_0$, and
$\rho_{2,{3/2}} = 1.8489$ $a_0$. The cutoff parameters for $\ell \ge 3$
were set to a common values of 1.897 $a_0$.
Table \ref{tab1} gives the calculated B-spline and experimental energies
coming from~\cite{nistasd500}.  The calculations with the core-polarization
potential are identified as the Dirac-Fock plus core polarization (DFCP)
model.  Differences between DFCP and experimental energies mostly occur in
the fourth digit after the decimal point.

One of the interesting aspects of Table \ref{tab1} concerns the spin-orbit
splitting of the $4p_J$ and $5p_J$ states.  The polarization potential
parameters $\rho_{1,1/2}$ and $\rho_{1,3/2}$ were tuned to give the correct
spin-orbit splitting of the $4p_J$ states.  Making this choice resulted in
the spin-orbit splittings for the $5p_J$ states also being
very close to experiment.

\begin{table}
\caption{Comparison of the electric dipole (E1),
electric quadrupole (E2) reduced matrix elements of several
interested states of the Ca$^{+}$ ion.  }
\label{tab3}
\begin{ruledtabular}
\begin{tabular}{lccc}
\multicolumn{1}{l}{Transition} &\multicolumn{1}{c}{DFCP}
&\multicolumn{1}{c}{MBPT-SD} &\multicolumn{1}{c}{RCC} \\
\hline
\multicolumn{4}{c}{Dipole} \\
$4s_{1/2}-4p_{1/2}$    & 2.879    &2.898(13) ~\cite{safronova11a}  & 2.88(1) \cite{sahoo09a}   \\
$4s_{1/2}-4p_{3/2}$    & 4.073    &4.099(18) ~\cite{safronova11a}  & 4.03(1) \cite{sahoo09a}   \\
$4s_{1/2}-5p_{1/2}$    & 0.089    &                                                           \\
$4s_{1/2}-5p_{3/2}$    & 0.109    &                                                           \\
$3d_{3/2}-4p_{1/2}$    & 2.500    &2.464(16) ~\cite{safronova11a}   &2.40(2) \cite{sahoo09a}   \\
$3d_{3/2}-4p_{3/2}$    & 1.116    &1.100(6) ~\cite{safronova11a}   &1.09(1) \cite{sahoo09a}   \\
$3d_{5/2}-4p_{3/2}$    & 3.356    &3.306(18) ~\cite{safronova11a}   &3.22(4) \cite{sahoo09a}   \\
$3d_{3/2}-5p_{1/2}$    & 0.091    &                   &                          \\
$3d_{3/2}-5p_{3/2}$    & 0.044    &                   &                          \\
$3d_{5/2}-5p_{3/2}$    & 0.131    &       \\
$3d_{3/2}-4f_{5/2}$    & 1.964    &1.927(52) \cite{safronova11a}                              \\
$3d_{5/2}-4f_{5/2}$    & 0.526    &0.516(6) \cite{safronova11a}                              \\
$3d_{5/2}-4f_{7/2}$    & 2.354    &2.309(29) \cite{safronova11a}                              \\
$4p_{1/2}-5s_{1/2}$    & 2.081    &2.073(11) \cite{safronova11a} &                       \\
$4p_{1/2}-3d_{3/2}$    & 4.205    & 4.28(3) \cite{safronova11a} &                       \\
$4p_{3/2}-4d_{3/2}$    & 1.894    & 1.93(1) \cite{safronova11a} &                       \\
$4p_{3/2}-4d_{5/2}$    & 5.675    & 5.78(3) \cite{safronova11a} &                       \\
\multicolumn{4}{c}{Quadrupole} \\
${4s_{1/2}}-3d_{3/2}$   & 8.120    &7.939(37) \cite{kreuter05a}  &  7.973 \cite{sahoo06a} \\
                                   &          &                  &  8.12(5) \cite{arora12a}   \\
${4s_{1/2}}-3d_{5/2}$    & 9.964   &9.740(47) \cite{kreuter05a}  & 9.979 \cite{sahoo06a}   \\
                                             &         &        &  9.97(6) \cite{arora12a}   \\
\end{tabular}
\end{ruledtabular}
\end{table}

\section{Transition matrix elements and associated quantities}

\subsection{Reduced Matrix Elements}

The dipole matrix elements were computed with a modified transition operator
\cite{hameed68a,hameed72a,mitroy88d},
e.g.
\begin{equation}
r{\bf C}^{1} = r{\bf C}^{1} - \left(1 - \exp(-r^6/\rho^6) \right)^{1/2} \frac{\alpha_{\rm core} r{\bf C}^{1}}{r^3}
\label{dipole}
\end{equation}
The cutoff parameter, $\rho$ used in Eq.~(\ref{dipole}) was set to
$\rho = (\rho_{\ell_a,j_a} + \rho_{\ell_b,j_b})/2$ where $a,b$ refer
to the initial and final states of the transition.

The static quadrupole polarizability of the Ca$^{2+}$ core is needed
for the calculation of the lifetimes of the
$3d_J$ states.  It was set $\alpha_{q,{\rm core}}= 6.936$ a.u.
\cite{johnson83a}.

There have been a number of previous calculations of reduced matrix elements
and polarizabilities for the low-lying states of Ca$^+$.  The semi-empirical
configuration interaction plus core polarization (CICP) can be regarded
as a non-relativistic predecessor of the present calculation
\cite{mitroy88d,mitroy08b}. Another method used is
the relativistic all-order single-double method where all single and double
excitations of the Dirac-Fock (DF) wave function are included to all orders
of many-body perturbation theory (MBPT-SD) \cite{safronova98a,arora07a,safronova11a}.  There have also been calculations using the relativistic coupled cluster
(RCC) method \cite{sahoo09a}.  The RCC and MBPT-SD approaches have many
common features \cite{pal07a,safronova08a,porsev12a}.
Atomic parameters computed using the RCC
approach have on a number of occasions had significant
differences with independent calculations
\cite{wansbeek08a,wansbeek10a,safronova11a,mitroy08d}.

\begin{table}
\caption{Comparison of the line strengths ratios for transitions involving
various spin-orbit doublets.   The notation $4s_{1/2}-4p_{3/2:1/2}$  means
the line strength ratio defined by dividing $4p_{3/2}$ line strength by
the $4p_{1/2}$ line strength. }
\label{tab4}
\begin{ruledtabular}
\begin{tabular}{lccc}
\multicolumn{1}{c}{Transition} &\multicolumn{1}{c}{DFCP} &\multicolumn{1}{c}{MBPT-SD} &\multicolumn{1}{c}{RCC} \\
\hline
$4s_{1/2}-4p_{3/2:1/2}$  &   2.0014  & 2.001~\cite{safronova11a}   & 1.958(17)~\cite{sahoo09a}        \\
$4s_{1/2}-5p_{3/2:1/2}$  &   1.4990  &       &      \\
$3d_{3/2}-4p_{3/2:1/2}$  &   5.0182  & 5.02~\cite{safronova11a}    & 4.85(12)~\cite{sahoo09a}    \\
$4p_{3/2}-3d_{5/2:3/2}$  &   9.043   & 9.04~\cite{safronova11a}    & 8.73(27)~\cite{sahoo09a}    \\
$4s_{1/2}-3d_{5/2:3/2}$  &   1.5057  & 1.5052~\cite{safronova11a}  & 1.5665~\cite{sahoo06a} \\
                         &           &                             & 1.5075~\cite{arora12a}\\
$4p_{3/2}-4d_{5/2:3/2}$  &   5.0181  &           \\
$4p_{3/2}-4d_{5/2}$      &   9.021   &  \\
\end{tabular}
\end{ruledtabular}
\end{table}

The reduced matrix elements between the various low lying states are the dominant
contributor to the polarizabilities of the $4s$, $3d$ and $4p$ levels.  These are
given in Table \ref{tab3} and compared with the results from other recent
calculations.  The ratio of line strengths for spin-orbit doublets is also interesting
to tabulate since they can reveal the extent to which dynamical effects (as opposed
to geometric effects caused by the different angular momenta) are affecting the
matrix elements.  Some line strength ratios are given in Table \ref{tab4}.

The variation between the DFCP, MBPT-SD and RCC matrix elements listed in
Table \ref{tab4} does not exceed 5$\%$.  The DFCP matrix elements are
usually closer to the MBPT-SD calculations than the RCC matrix elements.
A better indication of the differences between the DFCP, MBPT-SD and RCC
calculations is gained by examination of the line strength ratios listed
in Table \ref{tab4}.   The DFCP line strength ratios are within 1$\%$ of
the values that would be expected simply due to the
angular momentum factors alone.  The ratios are in very good agreement
with the MBPT-SD ratios.  It should be noted, that the line strength
ratios for the resonant transition of potassium have been measured to be
very close to 2.0 \cite{holmgren12a} and DFCP and MBPT-SD calculations also
predict line strength ratios very close to 2.0 \cite{holmgren12a,jiang13a}.

By way of contrast, RCC matrix element ratios exhibit about 4$\%$ differences
from the geometric ratios.  One would expect the RCC matrix element ratios
to be much closer to the MBPT-SD ratio given the close formal similarities
between the RCC and MBPT-SD approaches.  The RCC matrix element ratios listed
in Table
\ref{tab4} also show significant differences from the geometric ratio for
the $4s \to 3d_{J}$ transitions.    The DFCP and MBPT-SD ratios lie within
1$\%$ of the geometric ratios.  It should be noted that a similar situation
exists for the $5s - 4d_{5/2:3/2}$ line strength ratios of Sr$^+$
with RCC calculations exhibiting much larger differences due to non-geometric
effects than other calculations \cite{mitroy08c}.  The feature
common to the DFCP and MBPT-SD methods is that they use large B-spline
basis sets and calculated quantities are expected to be independent of basis
set effects.  One possible cause for the different RCC matrix element ratios
lies in the gaussian basis set used to represent virtual excitations in the
RCC calculation.  This point will be addressed later where polarizabilities
are discussed.

\begin{table}[th]
\caption{Lifetime of the $3d_{3/2}$ and $3d_{5/2}$ levels of Ca$^+$ (in
sec). The $3d_{3/2}:3d_{5/2}$ lifetime ratio is also given.
}
\label{tab5}
\begin{ruledtabular}
\begin{tabular}{lccc}
Source                   &   $\tau_{3d_{3/2}}$  & $\tau_{3d_{5/2}}$    & Ratio \\ \hline
DFCP                          & 1.143(1)(s)  & 1.114(1)  & 1.0260   \\
MBPT-SD \cite{safronova11a}   & 1.196(1)(s)  & 1.165(11) & 1.0266   \\
RCC \cite{sahoo09a}           & 1.185(7)     & 1.110(9)  & 1.0675     \\
MCHF \cite{vaeck92a}          & 1.160        & 1.140     & 1.0175     \\
Experiment \cite{kreuter05a}  & 1.176(11)    & 1.168(7)  & 1.007(15)  \\
Experiment \cite{lidberg99a}  & 1.17(5)      & 1.09(5)   & 1.073(90)  \\
Experiment \cite{gudjons96a}  & 1.064(17)    &           &            \\
Experiment \cite{knoop95a}    & 1.111(46)    & 0.994(38) & 1.118(80)  \\
\end{tabular}
\end{ruledtabular}
\end{table}

\subsection{Lifetimes}

The two most important lifetimes for the Ca$^+$ clock
\cite{matsubura08a,chwalla09a,hunag11b,huang12a,matsubura12a}
are the lifetimes
of the $3d_J$ and $4p_J$ levels.  The $3d_J$ states decay to the ground
state in an electric quadrupole transition with lifetime of about
1.1 sec \cite{kreuter05a}.  The $4p_J$ states experience electric
dipole transitions to both the $3d_J$ and $4s$ states.  Table \ref{tab5}
gives the lifetimes of the $3d_J$ states while Table \ref{tab6}
gives the lifetimes of the $4p_J$ states.  All DFCP lifetimes were
computed using experimental energy differences.

The most recent experiment for the $3d_J$ lifetimes give a ratio of
$1.007 \pm 0.015$ sec for the $3d_{3/2}$ and $3d_{5/2}$ states.  This
suggests that the $4s \to 3d_J$ matrix element ratios should be
close to the values expected from angular momentum coupling
considerations.  Older experiments \cite{knoop95a,lidberg99a} give
ratios further from unity, but in these cases the uncertainties are
much larger.

\begin{table}[th]
\centering \caption{Lifetimes (in nsec.) of the
$4p_{\frac{1}{2}}$ and $4p_{\frac{3}{2}}$ states.
The $4p_{\frac{1}{2}}:4p_{\frac{3}{2}}$ lifetime
ratio is also given.
The quantity $R$ gives fraction of the total decay rate
arising from the indicated transition.
}
\label{tab6}
\begin{ruledtabular}
\begin{tabular}{lllllll}
Level                 & DFCP            &MBPT-SD   &  RCC   & Expt.  \\
                      &                 &\cite{safronova11a}   & \cite{sahoo09a}   &   \\
\hline
$4p_{\frac{1}{2}}(ns)$     &6.94(1)   &6.88(6)    &6.931          & 7.098(20) \cite{jin93a}  \\
                           &            &           &               & 7.07(7) \cite{gosselin88a}  \\
$4p_{\frac{3}{2}}(ns)$     &6.75(1)   &6.69(6)    &6.881          & 6.926(19) \cite{jin93a}  \\
                           &            &           &               & 6.87(6) \cite{gosselin88a}  \\
                           &            &           &               & 6.72(2) \cite{gallagher67a}  \\
                           &            &           &               & 6.61(30) \cite{rainbow76a}  \\
$R(4p_{\frac{1}{2}}-4s_{\frac{1}{2}})$  &0.9324      &           &0.9374(74)     &           \\
$R(4p_{\frac{1}{2}}-3d_{\frac{3}{2}})$  &0.0676      &           &0.0626(5)      &           \\
$R(4p_{\frac{3}{2}}-4s_{\frac{1}{2}})$  &0.9313      &0.9340     &0.9350(62)     &0.9347(3) \cite{gerritsma08a}  \\
$R(4p_{\frac{3}{2}}-3d_{\frac{3}{2}})$  &0.0069      &0.00667    &0.00666(4)     &0.00661(4) \cite{gerritsma08a} \\
$R(4p_{\frac{3}{2}}-3d_{\frac{5}{2}})$  &0.0617      &0.0593     &0.0583(4)      &0.0587(2)  \cite{gerritsma08a} \\
Ratio                      &1.0281      &1.0284     &1.0073         &1.025(3)   \cite{jin93a}   \\
                                        &            &           &               &1.029(14)  \cite{gosselin88a}  \\
\end{tabular}
\end{ruledtabular}
\end{table}

The lifetimes of the $4p_J$ states depend on two transitions, these
are the $4s$-$4p_J$ and $3d_{J'}$-$4p_{J}$ transitions, with the
$4s$-$4p_J$ transition being the most important.   The lifetimes
and branching ratios for the $4p_J$ states are given in Table \ref{tab6}.
It not possible to reconcile the theoretical and experimental lifetimes
at the 1$\%$ level.  The two most recent experiments \cite{gosselin88a,jin93a} gave
lifetimes that are 2$\%$ larger than the DFCP lifetimes and 3$\%$ larger
than the MBPT-SD lifetimes.  Older Hanle effect experiments
\cite{gallagher67a,rainbow76a} gave lifetimes
closer to the MBPT-SD and DFCP lifetimes.

Measurements of the branching
ratios of the $4p_{3/2}$ state yield a picture where the MBPT-SD calculations
largely agree with experiment while the DFCP tends to overestimate
the contributions of the decays to the $3d_{J}$ levels.  Another area
of partial agreement between theory and experimental occurs for the
$4p_{1/2}:4p_{3/2}$ lifetime ratio.  The DFCP, MBPT-SD and experimental
ratios range from 1.025 to 1.030, with the RCC calculation again providing
an outlier at 1.0073.

\section{Polarizabilities}

\begin{table*}
\caption{Dipole and quadrupole polarizabilities (in a.u.) for low-lying states of the Ca$^{+}$ ion.
Non-relativistic quadrupole polarizabilities are not given for states with $\ell > 0$. The
RCC-STO results are those from Ref.~\cite{sahoo09a} that used a Slater type orbital basis
to represent virtual excitations.   }
\label{tab7}
\begin{ruledtabular}
\begin{tabular}{lcccccc}
&\multicolumn{2}{c}{$\alpha^{(0)}_{1}$}&\multicolumn{2}{c}{$\alpha^{(t)}_{1}$}
&\multicolumn{2}{c}{$\alpha_2$}   \\
\cline{2-3} \cline{4-5} \cline{6-7}
\multicolumn{1}{l}{State}&\multicolumn{1}{c}{DFCP}&\multicolumn{1}{c}{Others}
&\multicolumn{1}{c}{DFCP}&\multicolumn{1}{c}{Others}
&\multicolumn{1}{c}{DFCP}&\multicolumn{1}{c}{Others}\\
\hline
$4s_{1/2}$ & 75.28 &76.1(5)~MBPT-SD \cite{safronova11a} &          &                     & 882.43   & 871(4)~MBPT-SD \cite{safronova11a}  \\
           &       &75.49~CICP \cite{mitroy08b}      &          &                        &      & 875.1~CICP \cite{mitroy08b}                      \\
           &       &73.0(1.5)~RCC \cite{sahoo09a}   &          &                         &       & 712.9(24)~RCC \cite{sahoo07a}            \\
           &       &75.3(4)~$f$-sums \cite{chang83a}  &          &                           &       & 906(5) RCC \cite{arora12a}          \\
           &       &74.3~RCC-STO \cite{sahoo09a}   &          &                         &       &             \\
$4p_{1/2}$ & $-$2.774 &$-$0.75(70)~MBPT-SD \cite{safronova11a}&          &                                &7.466[4]  &     \\
           &       &-2.032~CICP \cite{mitroy08b}     &          &                                &        &          \\
$4p_{3/2}$ & $-$0.931 &1.02(64)~MBPT-SD \cite{safronova11a}&10.12    &10.31(28)~MBPT-SD \cite{safronova11a}   &$-$3.571[4] &        \\
           &       &-2.032~CICP \cite{mitroy08b}     &          &10.47~CICP \cite{mitroy08b}          &          &           \\
3$d_{3/2}$ &32.99  &32.0(3)~MBPT-SD \cite{safronova11a} &$-$17.88  &$-$17.43(23)~MBPT-SD \cite{safronova11a} &4928   &  \\
           &       &32.73~CICP \cite{mitroy08b}      &          &$-$17.64~CICP~\cite{mitroy08b}       &           &  \\
           &       &28.5(1.0)~RCC \cite{sahoo09a}   &          & $-$15.87~RCC~\cite{sahoo09a}      &           & \\

           &      &31.6~RCC-STO \cite{sahoo09a}   &          &$-$17.7~RCC-STO \cite{sahoo09a}     &       &      \\
3$d_{5/2}$ &32.81  &31.8(3)~MBPT-SD \cite{safronova11a} &$-$25.16  &$-$24.51(29)~MBPT-SD \cite{safronova11a} &$-$3304 & $-$3706(75) RCC \cite{arora12a}      \\
           &       &32.73~CICP \cite{mitroy08b}      &          &$-$25.20~CICP \cite{mitroy08b}       &       &                                         \\
           &      &29.5(1.0)~RCC \cite{sahoo09a}   &          &$-$22.49(5)~RCC \cite{sahoo09a}     &       &      \\
           &      &32.5~RCC-STO \cite{sahoo09a}   &          &$-$25.5~RCC-STO \cite{sahoo09a}     &       &      \\
\end{tabular}
\end{ruledtabular}
\end{table*}

\subsection{Static Polarizabilities}

The static dipole and quadrupole polarizabilities are calculated
by the usual sum-rule
\begin{equation}
\alpha_{(\ell)} = \sum_i \frac{ f^{(\ell)}_{gi} } {\varepsilon_{gi}^2}
\end{equation}
where the $f^{(\ell)}_{gi}$ are the absorption oscillator
strengths and $\varepsilon_{gi}$ is the excitation
energy of the transition.   Static dipole polarizabilities
for the $4s$, $4p_J$ and $3d_J$ states are listed in
Table \ref{tab7}.  All polarizabilities were computed
using experimental energy differences.

The most important polarizability is that of the $4s$ ground state
and there is only a 1$\%$ variation between the DFCP,
MBPT-SD and CICP static dipole polarizabilities.  The DFCP
polarizability is smaller than the MBPT-SD polarizability
because the DFCP $4s-4p_J$ matrix elements are smaller.
The RCC calculation of the dipole polarizability is the
clear outlier at 73.0 a.u. \cite{sahoo09a}.  The good agreement
between the DFCP, CICP and MBPT-SD polarizabilities does not
necessarily imply a 1$\%$ reliability in these polarizabilities
since the calculations give lifetimes for the $4p_J$ states that
are 2-3$\%$ smaller than experiment.

The variation between the DFCP, MBPT-SD and CICP estimates of
the $3d_J$ state polarizabilities do not exceed 1.0 a.u.
The difference in the polarizabilities for the two members
of the spin-orbit doublet is only 0.2 a.u.

The polarizabilities of the $4p_J$ states are close to zero with
the polarizability of the $4p_{3/2}$ state being about 1.8 a.u.
larger than the polarizability of the $4p_{1/2}$ state.
The polarizability is small because the downward transitions to
the  $4s_{1/2}$ and $3d_J$ states have negative oscillator strengths
which result in cancellations in the oscillator strength sum.  This
is evident in Tables \ref{D1} and \ref{D2} which show the breakdown
of the different contributions to the polarizabilities from the oscillator
strength sum rule.

The comparisons of the polarizabilities suggest that the basis
set used in the RCC calculations \cite{sahoo09a} could be improved.  The
recommended results for the RCC calculation are those computed with the
gaussian basis.  However, RCC calculations performed using a
Slater type orbital basis \cite{sahoo09a} give polarizabilities that
are in much better agreement with the MBPT-SD and DFCP polarizabilities.

\begin{table}[th]
\caption{The contributions of individual transitions
to the polarizabilities of the $4s_{1/2}$ and $4p_{1/2}$ states
at the magic wavelengths.  The numbers in brackets are uncertainties
in the last digits of the energy or wavelength calculated by introducing
2$\%$ uncertainties into the most important matrix elements.
}
\label{D1}
\begin{ruledtabular}
\begin{tabular}{lcccc}
$\omega$ (a.u.)  & 0         & 0.0659561(11247)    & 0.1152981(4)     & 0.1238091(303)   \\
$\lambda$ (nm)  & $\infty$   &690.817(11.984)     & 395.1807(14)     &368.0149(901)  \\
\hline
\multicolumn{5}{c}{$4s_{1/2}$}    \\
$4p_{1/2}$      & 24.0704    & 35.9364     &$-$2665.2940  &$-$147.2228  \\
$5p_{1/2}$      &  0.0097    &  0.0102     &    0.0117    &     0.0121    \\
$4p_{3/2}$      & 47.7532    & 70.6856     & 5558.6017    &$-$333.5265    \\
$5p_{3/2}$      &  0.0145    &  0.0153     &    0.0175    &     0.0181    \\
Remainder       &  0.1672    &  0.1710     &    0.1794    &     0.1815  \\
Core            &  3.2600    &  3.2664     &    3.2793    &     3.2823    \\
Total           & 75.2751    &110.0849     & 2896.7954    &$-$477.2554   \\
\multicolumn{5}{c}{$4p_{1/2}$}    \\
$4s_{1/2}$      &$-$24.0704  &$-$35.9364   &2665.2940     &   147.2228   \\
$5s_{1/2}$      &11.7449     & 16.4949     &  97.8655     &$-$798.9861    \\
$3d_{3/2}$      &$-$39.6152  & 69.1092     &  10.4051     &     8.7196    \\
$4d_{3/2}$      &40.8730     & 51.6866     & 113.3267     &   155.5319     \\
Remainder       & 5.0332     &  5.4542     &   6.6245     &     6.9740  \\
Core            & 3.2600     &  3.2664     &   3.2793     &     3.2823       \\
Total           & $-$2.7742  &110.0850     &2896.7954     &$-$477.2554 \\
\end{tabular}
\end{ruledtabular}
\end{table}

\begin{figure}[tbh]
\centering{
\includegraphics[width=8.5cm,angle=0]{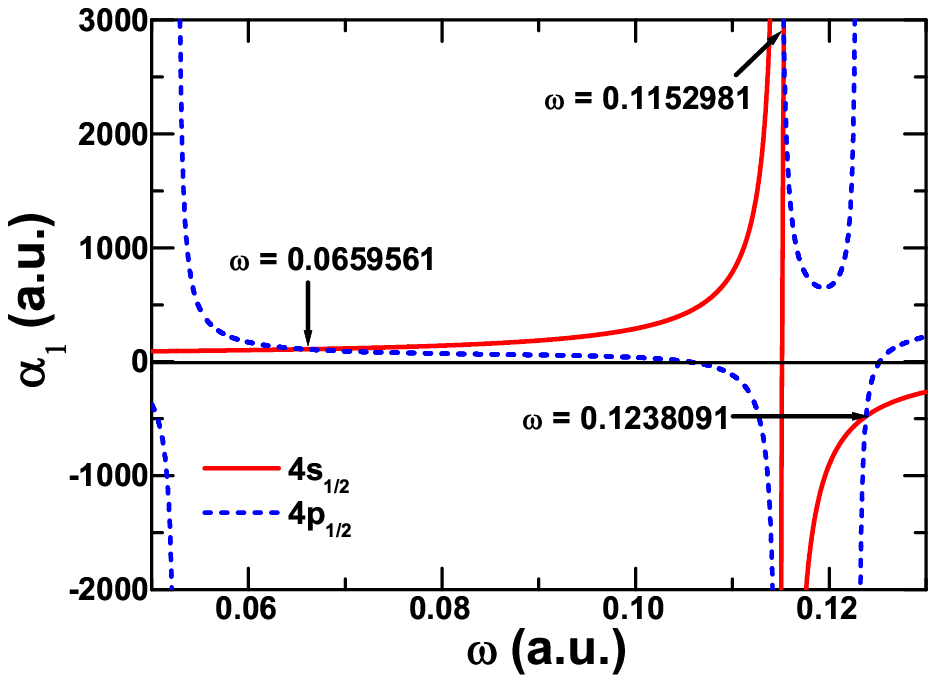}
}
\caption[]{
\label{fig1}
(color online) Dynamic polarizabilities of the $4s_{1/2}$ and $4p_{1/2}$ states of the
Ca$^+$ ions.  Magic wavelengths are identified by arrows.
}
\end{figure}

\begin{figure}[tbh]
\centering{
\includegraphics[width=8.5cm,angle=0]{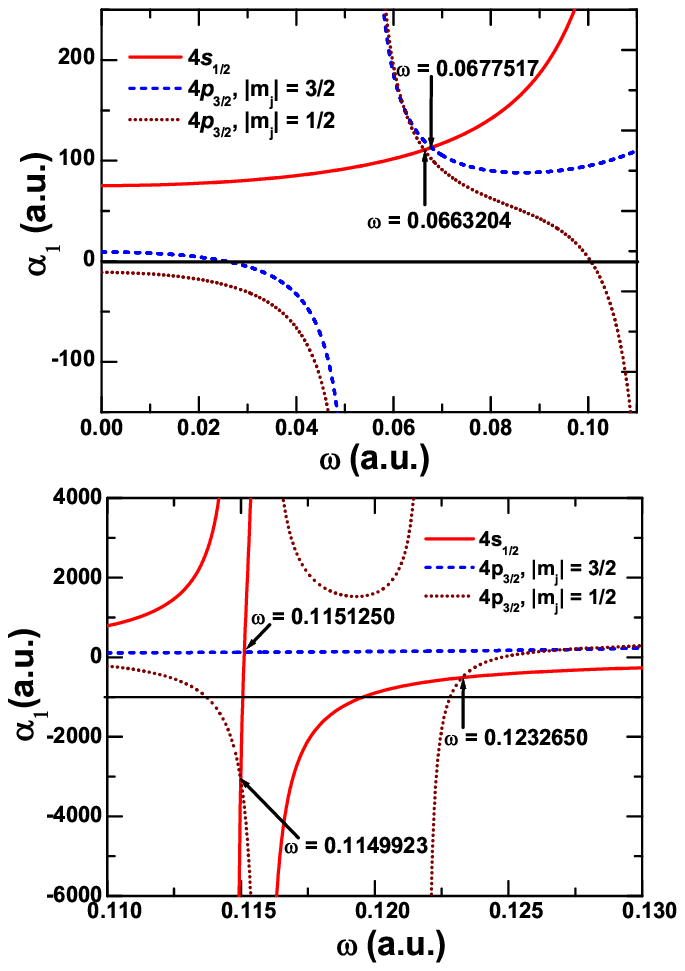}
}
\caption[]{
\label{fig2}
(color online) Dynamic polarizabilities of the $4s_{1/2}$ and $4p_{3/2}$ states of Ca$^+$.
Magic wavelengths are identified by arrows.
}
\end{figure}

\begin{table*}[th]
\caption{The contributions of individual transitions
to the polarizabilities of the $4s_{1/2}$ and $4p_{3/2}$ states
at the magic wavelengths.  These results assume non-polarized light.
The numbers in brackets are uncertainties in the last digits
calculated by assuming certain matrix elements have $\pm2\%$ uncertainties.
}
\label{D2}
\begin{ruledtabular}
\begin{tabular}{lccccccccccccccc}
$\omega$ (a.u.)& 0         & 0.0663204(11651)& 0.1149923(4) & 0.1232650(511) & 0.0677517(11210)   & 0.1151251(3)    \\
$\lambda$ (nm)& $\infty$  & 687.022(12.285) & 396.2315(13)&369.6393(1534)  & 672.508(11.3150)  & 395.7748(10)       \\
\hline
&\multicolumn{6}{c}{$4s_{1/2}$}    \\
$4p_{1/2}$      & 24.0704    & 36.1337    &$-$6530.5659    &$-$157.0218   & 36.9414    &$-$4009.0830     \\
$5p_{1/2}$      &  0.0097    &  0.0103    &      0.0117    &     0.0121   &  0.0103    &      0.0117     \\
$4p_{3/2}$      & 47.7532    & 71.0638    &   3449.6093    &$-$358.6372   & 72.6100    &   4129.0858     \\
$5p_{3/2}$      &  0.0145    &  0.0154    &      0.0175    &     0.0181   &  0.0154    &      0.0175     \\
Remainder       &  0.1672    &  0.1710    &      0.1794    &     0.1813   &  0.1712    &      0.1794     \\
Core            &  3.2600    &  3.2664    &      3.2791    &     3.2820   &  3.2667    &      3.2792     \\
Total           & 75.2751    &110.6606    &$-$3077.3881    &$-$512.1655   &113.0150    &    123.4906     \\
 &    \multicolumn{6}{c}{$4p_{3/2}$}\\
 &                   Average     & $m_{j}=1/2$& $m_{j}=1/2$ & $m_{j}=1/2$ & $m_{j}=3/2$  & $m_{j}=3/2$ \\
$4s_{1/2}$      &$-$11.9383  &$-$71.0636  &$-$3449.6902    &   358.6371   &   0.0000   &      0.0000      \\
$5s_{1/2}$      &    6.0501  &   34.3769  &    219.9519    &$-$1069.3049  &   0.0000   &      0.0000      \\
$3d_{3/2}$      &$-$ 5.4283  &    1.4608  &      0.2153    &     0.1808   &  11.6778   &      1.9317      \\
$4d_{3/2}$      &    5.8429  &    1.0626  &      2.3512    &     3.2245   &   9.6796   &     21.2507    \\
$3d_{5/2}$      &$-$31.6965  &   77.2921  &     11.5779    &     9.7303   &  45.8671   &      7.6960     \\
$4d_{5/2}$      &   33.7190  &   57.2196  &    126.3638    &   173.0763   &  38.6057   &     84.5964     \\
Remainder       &    4.3193  &    7.0456  &      8.5623    &     9.0081   &   3.9179   &      4.7365     \\
Core            &    3.2600  &    3.2664  &      3.2791    &     3.2820   &   3.2667   &      3.2792     \\
Total           &$-$ 4.1279  &  110.6606  &$-$3077.3881    &$-$512.1655   & 113.0150   &    123.4906     \\

\end{tabular}
\end{ruledtabular}
\end{table*}

\begin{table}[th]
\caption{Pseudo-spectral oscillator strength distribution used
in the computation of the dynamic polarizability of the Ca$^{2+}$ core.
Energies are given in a.u..  }
\label{tab8}
\begin{ruledtabular}
\begin {tabular}{lcc}
$i$ &    $\varepsilon_i$   & $f_i$  \\
\hline
1 &   133.689002  &    2.0  \\
2 &    14.645933  &    2.0  \\
3 &    11.675258  &    6.0  \\
4 &    1.9047772  &    2.0  \\
5 &    1.1104171  &    6.0  \\
\end{tabular}
\end{ruledtabular}
\end{table}

\begin{table*}[H]
\centering \caption{ The contributions of individual transitions
to the polarizabilities of the $4s_{1/2}$ and $3d_{5/2}$ states
at the magic wavelengths.  These results assume non-polarized light.
The numbers in brackets are uncertainties in the last digits
calculated by assuming certain matrix elements have $\pm2\%$ uncertainties
as described in the text.
}
\label{D3}
\begin{ruledtabular}
\begin{tabular}{lcccccc}
$\omega$ (a.u.)    &$0$          & 0.0340414(22387) & 0.0424109(10654) & 0.1151182(1)  & 0.1151184(1)  & 0.1151186(1)   \\
$\lambda$(nm)      &$\infty$     & 1338.474(82.593) & $1074.336(26.352)$  &$395.7982(1)  $ &$395.7978(1)$  & $395.7968(1)$  \\
\hline
  &  \multicolumn{6}{c}{$4s_{1/2}$}\\
${4p_{1/2}}$         &24.0704    &26.3917     &27.8762     &$-$4090.5249    &$-$4088.7574   &$-$4085.2247       \\
${5p_{1/2}}$         & 0.0097    & 0.0098     & 0.0099     &      0.0117    &      0.0117   &      0.0117    \\
${4p_{3/2}}$         &47.7532    &52.2705     &55.1513     &   4087.5752    &   4088.4488   &   4090.2003     \\
${5p_{3/2}}$         & 0.0145    & 0.0147     & 0.0148     &      0.0175    &      0.0175   &      0.0175    \\
Remainder            & 0.1672    & 0.1682     & 0.1688     &      0.1794    &      0.1794   &      0.1793    \\
Core                 & 3.2600    & 3.2618     & 3.2627     &      3.2792    &      3.2792   &      3.2792    \\
Total                &75.2751    &82.1167     &86.4837     &      0.5371    &      3.1792   &      8.4633    \\
 &    \multicolumn{6}{c}{$3d_{5/2}$}\\
 &                   Average     & $m_{j}=1/2$& $m_{j}=3/2$ & $m_{j}=1/2$ & $m_{j}=3/2$  & $m_{j}=5/2$ \\
${4p_{3/2}}$         &29.5834    &71.3309     &76.6749     &-11.5457        &$-$7.6971      &0.0000    \\
${5p_{3/2}}$         &0.0113     & 0.0165     & 0.0119     &  0.0227        &0.0151         &0.0000    \\
${4f_{5/2}}$         &0.0607     & 0.0109     & 0.0988     &  0.0136        &0.1223         &0.3398    \\
${5f_{5/2}}$         &0.0196     & 0.0035     & 0.0318     &  0.0041        &0.0367         &0.1018    \\
${4f_{7/2}}$         &2.5573     & 3.2582     & 2.7444     &  4.0780        &3.3983         &2.0391    \\
${5f_{7/2}}$         &0.8270     & 1.0479     & 0.8799     &  1.2223        &1.0186         &0.6112    \\
Remainder            &2.5979     & 3.1870     & 2.7803     &  3.4628        &3.0060         &2.0922    \\
Core                 &3.2600     & 3.2618     & 3.2627     &  3.2792        &3.2792         &3.2792    \\
Total                &38.5915    &82.1167     &86.4837     &  0.5371        &3.1792         &8.4633    \\
\end{tabular}
\end{ruledtabular}
\end{table*}

\subsection{Dynamic polarizabilities and magic wavelengths}

The dynamic dipole polarizability of a state at photon energy
$\omega$ is defined
\begin{equation}
\alpha_{1}(\omega) = \sum_i \frac{ f^{(1)}_{gi} } {\varepsilon_{gi}^2 - \omega^2}
\end{equation}
The dipole polarizability has a tensor component for states with
states with $J > 1/2$.  This can be written
\begin{eqnarray}
\alpha^{\rm T}_{1}(\omega) &=&   6\left( \frac{5J_g(2J_g-1)(2J_g+1)}{6(J_g+1)(2J_g+3)}  \right)^{1/2} \nonumber \\
    & \times & \sum_{J_i}  (-1)^{J_g+J_i}
  \left\{ \begin{array}{ccc}
   J_g & 1 & J_i \\
   1 & J_g & 2
   \end{array} \right\}
\frac{ f^{(1)}_{gi} } {\varepsilon_{gi}^2 - \omega^2}
\end{eqnarray}
The polarizability for a state with non-zero angular momentum $J$ depends on the magnetic projection $M_g$:
\begin{equation}
\alpha_{1,M_g} = \alpha_{1} + \alpha^{T}_1 \frac{3M_g^2-J_g(J_g+1)}{J_g(2J_g-1)}. \label{alphaM}
\end{equation}

The dynamic polarizabilities includes contributions from  the core which
is represented by a pseudo-oscillator strength distribution
\cite{margoliash78a,kumar85a,mitroy03f} which is
tabulated in Table \ref{tab8}.  The distribution is derived from
the single particle energies of a Hartree-Fock core.  Each separate
$(n,\ell)$ level is identified with one transition with a pseudo-oscillator
strength equal to the number of electrons in the shell.  The excitation
energy is set by adding a constant to the Koopmans energies and
adjusting the constant until the core polarizability from the
oscillator strength sum rule is equal to the known core polarizability
of 3.26 a.u.  The core polarizabilities of any two states effectively
cancel each other when the polarizability differences are computed.

\begin{figure}[tbh]
\centering{
\includegraphics[width=8.5cm,angle=0]{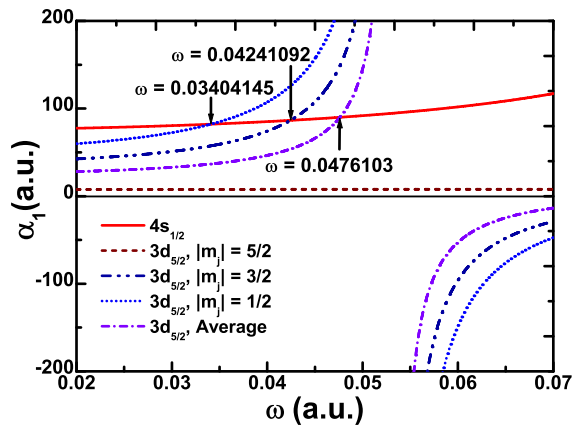}
}
\caption[]{ \label{fig3}
(color online)
Dynamic polarizabilities of the $4s_{1/2}$ and $3d_{5/2}$ states of Ca$^+$.
Magic wavelengths are identified by arrows.
}
\end{figure}

\begin{figure}[tbh]
\centering{
\includegraphics[width=8.5cm,angle=0]{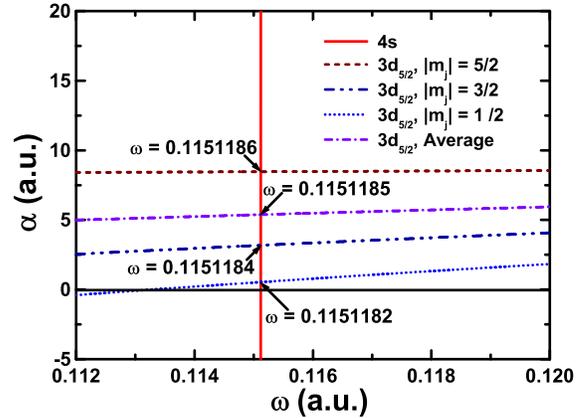}
}
\caption[]{ \label{fig4}
(color online) Dynamic polarizabilities of the $4s_{1/2}$ and $3d_{5/2}$ states of Ca$^+$.
Magic wavelengths is identified by circles and arrows.
}
\end{figure}

The dynamic polarizabilities for the $4s_{1/2}$ and $4p_{1/2}$ states of Ca$^+$
are shown in Figure \ref{fig1}.  The first magic wavelength occurs at
$\omega = 0.0659561$ a.u.  after the photon wavelength exceeds the energy for
the $4p_{1/2}$-$3d_{3/2}$ transition.
Magic wavelengths are identified at
$\lambda = 690.817$, 395.181 and 368.015 nm.  The 395.181 nm magic
wavelengths occur when the photon is very close to the excitation energies
of the $4s$-$4p_J$ states.  The 368.015 nm magic wavelength occurs
near the energy for the $4p_{1/2}-5s_{1/2}$ transition.
The dominant contributions to polarizabilities at the magic wavelengths
are listed in Table \ref{D1}.  The $4s$ polarizability
is dominated by the $4s_{1/2}$-$4p_{J}$ transitions with the next largest
contribution coming from the core.  However, the $4p_{1/2}$ polarizability has
significant contributions from the transitions to the $4s$, $5s$ and $3d_{3/2}$
states. A magic wavelength experiment would give information about the $4p_{1/2}$
state, but would not give detailed information about any individual matrix
element.  An experiment that measured all three magic wavelengths could conceivably
be able to extract information about individual line strengths, however it should
be noted that two of the transitions are in the ultraviolet.

\begin{table*}[H]
\caption{ The contributions of individual transitions
to the polarizabilities of the $4s_{1/2}$ and $3d_{3/2}$ states
at the magic wavelengths.  These results assume non-polarized light.
The numbers in brackets are uncertainties in the last digits
calculated by assuming certain matrix elements have $\pm2\%$ uncertainties
as described in the text.
}
\label{D4}
\begin{ruledtabular}
\begin{tabular}{llccccc}
 $\omega$            &0             & 0.0348188(20007) & 0.0513460(1855) & 0.0535831(1) & 0.1151182(1) & 0.1151185(1) \\
 $\lambda$           &$\infty$      & 1308.590(71.108) & 887.382(3.196)  & 850.335(2)  & 395.7981(1)  & 395.7970(1) \\
\hline
  & \multicolumn{6}{c}{$4s_{1/2}$} \\
${4p_{1/2}}$         & 24.0704      &26.5098     &30.0922     &30.7777      &$-$4090.1007  &$-$4086.1347\\
${5p_{1/2}}$         &  0.0097      & 0.0098     & 0.0100     & 0.0101      &      0.0117  &      0.0117\\
${4p_{3/2}}$         & 47.7532      &52.4999     &59.4402     &60.7642      &   4087.7840  &   4089.7488\\
${5p_{3/2}}$         &  0.0145      & 0.0147     & 0.0150     & 0.0151      &      0.0175  &      0.0175\\
Remainder            &  0.1672      & 0.1683     & 0.1695     & 0.1697      &      0.1794  &      0.1794\\
core                 &  3.2600      & 3.2619     & 3.2639     & 3.2643      &      3.2792  &      3.2792\\
Total                & 75.2751      & 82.4644    &92.9908     &95.0011      &      1.1711  &      7.1019\\
  &  \multicolumn{6}{c}{$3d_{3/2}$}\\
                     &Average       &$m_{j}=1/2$ &$m_{j}=3/2$ &$m_{j}=1/2$  &$m_{j}=1/2$   &$m_{j}=3/2$ \\
${4p_{1/2}}$         & 9.9038       & 70.5419     & 0          &$-$1034.8996 &$-$10.4461    &   0        \\
${5p_{1/2}}$         & 0.0033       &  0.0134     & 0          &      0.0139 &   0.0184     &   0        \\
${4p_{3/2}}$         & 5.4284       &  1.3416     &84.7097     &   1119.1728 &$-$0.2147     &$-$1.9320   \\
${5p_{3/2}}$         & 0.0021       &  0.0003     & 0.0029     &      0.0003 &   0.0004     &   0.0038   \\
${4f_{5/2}}$         & 2.3339       &  3.1745     & 2.1676     &      3.2644 &   3.9674     &   2.6449   \\
${5f_{5/2}}$         & 0.7556       &  1.0218     & 0.6928     &      1.0422 &   1.1908     &   0.7939   \\
Remainder            & 2.3601       &  3.1091     & 2.1539     &      3.1428 &   3.3757     &   2.3121   \\
core                 & 3.2600       &  3.2619     & 3.2639     &      3.2643 &   3.2792     &   3.2792   \\
Total                 &24.0472      & 82.4644     &92.9908     &     95.0011 &   1.1711     &   7.1019   \\
\end{tabular}
\end{ruledtabular}
\end{table*}

The dynamic polarizabilities of the $4s_{1/2}$ and $4p_{3/2}$ states of
Ca$^+$ are shown in Figure \ref{fig2}.  These figures assume non-polarized
light.  Figure \ref{fig2} only has two magic wavelengths  below
$\omega = 0.125$ a.u.  Transitions to the
$ns_{1/2}$ states make no contribution to the $4p_{3/2}$ state polarizability.
This is evident from Table \ref{D2} which details the breakdown of different
transitions to the polarizability.   The magic wavelength at 395.775 nm for
the $4p_{3/2,m=3/2}$ magnetic sub-level can give an estimate of the contribution
to the $np_{3/2}$ polarizability arising from excitations to the $nd_{J}$ levels.

The $4s_{1/2}$ and $3d_{5/2}$ polarizabilities are shown in Figures \ref{fig3}
and \ref{fig4}.  The $3d_{5/2,m}$ polarizabilities are shown for all magnetic
sub-levels and also for the average polarizability.  Magic wavelengths occur
when the photon
energy gets close to the excitation energies for the $3d_{5/2} \! \to \! 4p_{J}$
transitions and the $4s_{1/2} \! \to \! 4p_{J}$ transitions.    Figure \ref{fig3}
shows the $4s_{1/2}$ and $3d_{5/2}$ polarizabilities at photon energies
between 0.02 and 0.07 a.u.  Precise values of the magic wavelengths
and the breakdown of the polarizability into different components can be
found in Table \ref{D3}.

Two of the magnetic sub-levels have magic wavelengths at infrared frequencies,
namely $\lambda = 1338.474$ and 1074.336 nm.  The contributions to the
in $3d_{5/2}$ polarizability are dominated by the $3d_{5/2} \! \to \! 4p_{3/2}$
transition which constitutes about 88$\%$ of the polarizability.  The
measurement of these magic wavelengths provides a method to determine the
$f_{4s_{1/2} \to 4p_{J}}$ to $f_{3d_{5/2} \to 4p_{3/2}}$ oscillator
strength ratios.  Suppose all the remaining components of the $3d_{5/2}$
polarizability can only be estimated to an accuracy of 10$\%$.  The overall
net uncertainty in the remaining terms would be less than 1.5$\%$.

There are also an additional magic wavelengths that can potentially
be measured.  The $4s$ dynamic polarizability goes through zero as
the wavelength passes through energies needed to excite the
$4s \! \to \! 4p_{1/2}$ and  $4s \! \to \! 4p_{3/2}$ transitions.  Figure
\ref{fig4} shows the polarizabilities for the $4s$ and $3d_{5/2}$ at energies
near the $4s \to 4p_{J}$ excitation energies.
The $3d_{5/2}$ polarizabilities are typically small in magnitude in
this wavelength range.  The magic wavelength arises more from the
the cancellation of the $4p_{1/2}$ and $4p_{3/2}$ contributions to
the $4s$ dynamic polarizability than from the cancellation between
the $4s$ and $3d_{5/2}$ dynamic polarizabilities.  Measurement of the
magic wavelength here is in some respects in analogous to a measurement
of the longest tune-out wavelength for neutral potassium \cite{jiang13a}.
Zero field shift wavelengths measured in the spin-orbit energy gap of
the resonant transition are strongly dominated by the large and
opposite polarizability contributions of the two members of the spin-orbit
doublet \cite{tang13a,jiang13a}.   This makes it possible to accurately
determine the oscillator strength ratio, i.e.
$f_{4s \to 4p_{1/2}}$:$f_{4s\to4p_{3/2}}$,  of the two transitions
comprising the spin-orbit doublet.

\begin{figure}[tbh]
\centering{
\includegraphics[width=8.5cm,angle=0]{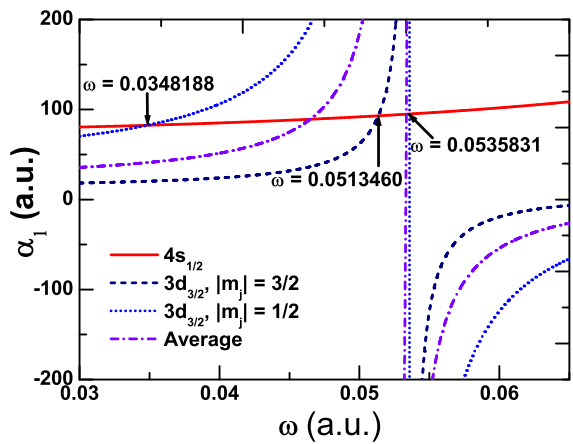}
}
\caption[]{ \label{fig5}
(color online) Dynamic polarizabilities of the $4s_{1/2}$ and $3d_{3/2}$ states of Ca$^+$.
Magic wavelengths are identified by arrows.
}
\end{figure}

Table \ref{D4} identifies the magic wavelengths associated with the
$4s \to 3d_{3/2}$ energy interval.  The situation here is similar
to the situation for the $4s \to 3d_{5/2}$ magic wavelengths.  However,
there are three  magic wavelengths in the infrared region of the spectrum.
This transition has an additional magic wavelength since the
$3d_{3/2,m=1/2}$ state, unlike the $3d_{5/2,m=1/2}$ state, also undergoes
undergoes a transition to the $4p_{1/2}$ state.  The polarizability
difference in the 0.02 to 0.07 a.u. energy range is plotted in Figure \ref{fig5}.
The $3d_{3/2}$ polarizability is dominated by the $3d_{3/2} \to 4p_J$
transition and a magic wavelength measurement can be used to make an
estimate of the $3d_{3/2} \to 4p_J$  line strength relative to the $4s$
dynamic polarizability.  The $3d_{3/2,m=1/2}$ polarizability at
850.335 nm has large contributions from the $4p_{1/2}$ and $4p_{3/2}$
states since it lies between the excitation energies of these of
states.  Measurement of the 850.335 nm and 1308.590 nm wavelengths
together would give estimates of the $3d_{3/2} \! \to \! 4p_{1/2}$ line
strengths and the $f_{3d_{3/2} \to 4p_{1/2}}$:$f_{3d_{3/2}\to4p_{3/2}}$
ratio.  A measurement of the magic wavelengths in the
vicinity 395 nm provides would permit a determination of the
$f_{4s \to 4p_{1/2}}$:$f_{4s\to4p_{3/2}}$ ratio.

\subsection{Uncertainties}

An uncertainty analysis has been done for all the magic wavelengths
presented in the preceding sections.  This analysis was aimed at
making an initial estimate of how uncertainties in the matrix elements
of the most important transitions would translate to a shift in
the magic wavelengths.  The primary purpose of the uncertainty
analysis is to define reasonable limits to help guide an
experimental search for the magic wavelengths identified in this
paper.

In the case of the $4s \! \to \! 4p_J$ polarizability differences, the
$4s \! \to \! 4p_J$, $4p_J \! \to \! 5s$, $4p_J \! \to \! 3d_J$ and $4p_J \to 4d_J$
matrix elements were all changed by $2\%$ and the magic wavelengths
recomputed.  The matrix elements involving the different spin-orbit
states of the same multiplet were all given the same scaling.
A variation of $\pm 2\%$ was chosen by reference to the
difference of the DFCP matrix elements with the experimental or
the MBPT-SD matrix elements.  The estimate of a $2\%$ uncertainty
in the $4s \to 4p_J$ matrix element can be regarded as a conservative
estimate.

The $4s \to 3d_J$ polarizability difference is predominantly
determined by the $4s \to 4p_J$ and $3d_J \to 4p_J$ matrix
elements.  So variations of $\pm 2\%$ in these two transitions
were used in determining the uncertainties in the magic wavelengths.

There are a number of magic wavelengths which are relatively insensitive
to changes in the matrix elements of a multiplet.  One of these
wavelengths is the 850 nm wavelength for the $4s-3d_{3/2}$ interval
and the others are the magic wavelengths near 395 nm.  These
wavelengths arise due to cancellations in the polarizabilities due
to two transitions of a spin-orbit doublet.  In the case of the
850 nm magic wavelength, the relevant transitions are the
$3d_{3/2} \! \to \! 4p_J$ transitions.

The sensitivity of the magic wavelengths near 395 nm to changes
in the transition matrix elements depends on the overall size of
the polarizabilities of the $4p_J$ and $3d_J$ levels.  When these
are large due to transitions other than the $4s \to 4p_J$ transition,
then the 395 nm magic wavelength shows higher sensitivity to the
changes in the matrix elements.  However, the net change in the
magic wavelengths for 2$\%$ changes in the matrix elements is
about 0.001 nm for the $4s \! \to \! 4p_J$ interval.  The sensitivity
to 2$\%$ matrix element changes for the $4s \! \to \! 3d_J$ intervals
is about 0.0001 nm due to the small polarizabilities of the
$3d_J$ states near 395 nm.   The 850 nm magic wavelength is also
relatively insensitive to changes in the overall size of the
matrix elements, with the 2$\%$ matrix element change leading
to a change of only 0.0001 nm in the magic wavelengths.
The low sensitivity of magic wavelengths to the
overall size of the matrix elements in these cases means that these
the magic wavelengths can be used to give precise estimates of the
matrix element ratios of the two transitions in the spin-orbit doublet.

The 1338, 1309, 1074, 887 nm magic wavelengths show much greater sensitivity
to 2$\%$ changes in the matrix elements.  The changes in the magic wavelengths
range from 3 to 80 nm.  The sensitivity of the magic wavelengths to these
matrix elements is driven by the rate of change of the $4s$ and $3d_J$
polarizabilities with energy.  A large change in the photon energy is needed to
compensate for a small change in the polarizability when $d\alpha_1/d\omega$ is small.
The sensitivity of the magic wavelength to small changes in the matrix elements
decreases as the photon energy gets closer to the $3d_J \! \to \! 4p_{J'}$ excitation
thresholds.   The high sensitivity of the magic wavelengths with respect to
changes in the matrix elements means it is only necessary to measure the
magic wavelength to a precision of 0.10 nm to impose reasonably tight constraints
on the ratios of the $4s \! \to \! 4p_J$ and $3d_J \! \to \! 4p_{J'}$ matrix element
rations.

\section{Conclusion}

A relativistic semi-empirical core model is applied to the calculation
of the dynamic polarizabilities of the $4s$, $3d_J$ and $4p_J$ states
of Ca$^+$.  A number of magic wavelengths at convenient photon energies
have been identified for the $4s$-$3d_J$ energy intervals.  Measurement of
these magic wavelengths can be used to determine reasonably accurate estimates
of the  $3d_J$-$4p_{J'}$ line strengths relative to the $4s$-$4p_J$
line strengths.  This could lead to improved estimates of the blackbody
radiation shift for the Ca$^+$ clock transition.  There is one
impediment.  At the moment there is a 3$\%$ spread between theoretical
and experimental lifetimes for the $4p_{J'}$ states.   This variation,
which does not exist for the same transition in potassium
\cite{safronova08b,jiang13a}, needs to resolved so the uncertainty
in the $4s-4p_{J}$ line strengths can be reduced to 1$\%$ or better.

There are two other relatively clean measurements of atomic structure
parameters that could be made.  Measurement of the magic wavelength near
395 nm could be used to determine a value of the oscillator strength
$f_{4s \to 4p_{1/2}}$:$f_{4s\to4p_{3/2}}$ ratio.  This could help resolve
the incompatible predictions of this ratio by DFCP/MBPT-SD and RCC
calculations.  Comparisons of polarizabilities do suggest that the
gaussian basis set used for the RCC calculations could be improved.
Further, measurements of the two longest magic wavelengths for the
$3d_{3/2,m=1/2} \to 4s_{1/2}$ transition could give a good estimate
of the $f_{3d_{3/2} \to 4p_{1/2}}$:$f_{3d_{3/2} \to 4p_{3/2}}$ ratio.

The utility of measuring magic wavelengths for selected Ca$^+$
transitions can of course be extended to other alkaline-earth
ions, with Sr$^+$ and Ba$^+$ being obvious possibilities.  A
single ion optical frequency standard at the 10$^{-17}$ level
of precision has recently been reported for the $5s$-$4d_{5/2}$
transition of the Sr$^+$ ion \cite{madej12a}.  It is likely
that the determination of the magic wavelengths for this
transition could be used to improve the precision of estimates
of the blackbody radiation shift for this transition
\cite{jiang09a,mitroy08c}.

\begin{acknowledgments}

This work was supported by the National Basis Research Program of China under Grant
Nos.2010CB832803 and 2012CB821305
and by NNSF of China under grant Nos. 11274348 and 11034009.
This research was supported by the Australian Research Council
Discovery Project DP-1092620.  We would like to thank Dr Jun Jiang of
CDU for assistance in manuscript preparation.

\end{acknowledgments}


\end{document}